\documentclass[sigconf]{acmart}
\pdfoutput=1
\usepackage{mdframed}
\usepackage{xcolor}
\usepackage[most]{tcolorbox} 
\usepackage{lipsum}
\usepackage{booktabs}   
\usepackage{multirow}   
\usepackage{array}      
\usepackage{arydshln}   
\usepackage{graphicx}   
\usepackage{caption}    
\usepackage{xparse}
\usepackage{booktabs}
\usepackage{makecell} 
\definecolor{findingsgray}{RGB}{245,245,245}

\AtBeginDocument{%
  }

\begin{document}

\title{LogPurge: Log Data Purification for Anomaly Detection via Rule-Enhanced Filtering}

\author{Shenglin Zhang$^{1}$, Ziang Chen$^{1}$, Zijing Que$^{1}$, Yilun Liu$^{2}$, Yongqian Sun$^{1,*}$, Sicheng Wei$^{1}$\\Dan Pei$^{3}$, Hailin Li$^{1}$}\thanks{$*$ Corresponding author.}
\affiliation{
\vspace{0.05cm}
\institution{$^{1}$ Nankai University \quad $^{2}$ Huawei \quad $^{3}$ Tsinghua University}
\country{}
}

\email{{zhangsl, sunyongqian}@nankai.edu.cn,{2120240792, 2120250759, 2212534, nkualexlee}@mail.nankai.edu.cn}
\email{liuyilun3@huawei.com, peidan@tsinghua.edu.cn}

\renewcommand{\shortauthors}{Shenglin Zhang, et al.}

\begin{abstract}
  Log anomaly detection, which is critical for identifying system failures and preempting security breaches, detects irregular patterns within large volumes of log data, and impacts domains such as service reliability, performance optimization, and database log analysis. Modern log anomaly detection methods rely on training deep learning models on clean, anomaly-free log sequences. However, obtaining such clean log data requires costly and tedious human labeling, and existing automatic cleaning methods fail to fully integrate the specific characteristics and actual semantics of logs in their purification process.
In this paper, we propose a cost-aware, rule-enhanced purification framework, LogPurge, that automatically selects a sufficient subset of normal log sequences from contamination log sequences to train a anomaly detection model. Our approach involves a two-stage filtering algorithm: In the first stage, we use a large language model (LLM) to remove clustered anomalous patterns and enhance system rules to improve LLM's understanding of system logs; in the second stage, we utilize a divide-and-conquer strategy that decomposes the remaining contaminated regions into smaller subproblems, allowing each to be effectively purified through the first stage procedure.
Our experiments, conducted on two public datasets and one industrial dataset, show that our method significantly removes an average of 98.74\% of anomalies while retaining 82.39\% of normal samples. Compared to the latest unsupervised log sample selection algorithms, our method achieves F-1 score improvements of 35.7\% and 84.11\% on the public datasets, and an impressive 149.72\% F-1 improvement on the private dataset, demonstrating the effectiveness of our approach.

\end{abstract}




\maketitle

\section{Introduction}
Log data is central to modern IT operations, enabling engineers to gain insights into the internal states of complex systems for monitoring, management, and troubleshooting~\cite{le2022log, zhang2024multivariate}. The effectiveness of log-based anomaly detection further impacts domains such as service reliability, performance optimization, and database log analysis\cite{zhang2025log,zhang2024multivariate}. However, as systems grow increasingly complex and log volumes explode, pinpointing issues and flagging critical anomalies within this deluge has become both crucial and challenging~\cite{le2024prelog}. To address this, researchers have proposed a series of unsupervised anomaly detection methods based on deep learning~\cite{du2017deeplog,guo2021logbert,nedelkoski2020self,sui2025bridging}. These approaches automatically learn feature representations of normal log sequences and then identify records deviating from this pattern as anomalies.

However, these approaches heavily rely on the assumption that the training data possess high data quality~\cite{le2022log,ma2024pluto}, meaning that all samples are normal and free from contamination. In practice, ensuring such clean data is both resource-intensive and labor-demanding. The presence of anomalous samples inevitably degrades the overall data quality, leading to a decline in model performance. Consequently, contamination has been shown to compromise the robustness of unsupervised methods (Section~\ref{RQ1:Effectiveness in Removing Contamination}). This raises a fundamental question: how can we obtain training data with high data quality that reliably preserves normal patterns while excluding anomalies?

\noindent \textbf{Log Purification Task.}
Given a contaminated log dataset containing both normal and abnormal log sequences,
the task is to extract a \emph{clean and representative subset} at the sequence level.
This subset should (i) retain sufficient normal sequences to capture the diversity of system behaviors,
and (ii) remove anomalous sequences that could corrupt the learned representations.
By constructing such a high-quality subset, unsupervised anomaly detection models can focus on normal data characteristics rather than being misled by abnormal log sequences.

Existing effort has only partially addressed this challenge. Ma~\emph{et al.} proposed \textsc{Pluto}~\cite{ma2024pluto} to distinguish high- and low-contamination regions in the embedding space by analyzing their statistical characteristics. For high-contamination regions, it removes clusters using a dominance-based metric. For low-contamination regions, it estimates the number of samples to be removed in each remaining cluster and filters those whose secondary feature vectors are highly aligned, which are likely to be anomalies.

However, this strategy still faces three major limitations: (1) \textbf{Semantic insufficiency in contamination assessment:} the metric for filtering high-contamination regions mainly depends on the concentration of data distributions rather than fully using semantic understanding of log sequences, whereas the regional concentration does not always directly correlate with the regional contamination;(2) \textbf{Coupling-induced estimation bias:} the strong coupling between high- and low-contamination regions makes the algorithm heavily rely on the correctness of their distinction to estimate the anomaly ratio of each region-an incorrect estimation can lead to bias reversal and cause the low-contamination filtering strategy to fail; and (3) \textbf{Impractical Hyperparameter:} the global anomaly ratio required for filtering low-contamination samples is difficult to obtain in practice, \emph{as we further discuss in Section~\ref{sec:limits-pluto}}.

Recently, large language models (LLMs) have demonstrated powerful semantic understanding in log analysis~\cite{liu2024logprompt,huang2025logrules,cui2025logeval} and data cleaning~\cite{sheng2025llms,liu2024coachlm}, offering new inspiration for addressing the above limitations of Pluto. Their ability to fully leverage the inherent semantics of logs enables LLMs to understand and reason about fine-grained log patterns~\cite{liu2025loglm,ji2025superlog}. This makes them suitable for interpretable analysis in log-related tasks. However, in our preliminary attempts, we observe two critical challenges that must be addressed before LLMs can be effectively applied to log purification:

(i) \textbf{The prohibitive inference cost} of applying LLMs to large-scale log purification tasks poses a significant challenge.
Directly employing LLMs to assess the anomaly level or purity of log sequences within each region requires extensive inference operations. As shown in Table~\ref{table: details of datasets}, the training datasets contain a large number of log sequences, making direct sequence-level judgment by LLMs computationally infeasible due to their substantial inference time and cost, as further demonstrated in ~\ref{RQ3:Cost-Effectiveness of LLM-based Purification for Anomaly Detection}.

(ii) \textbf{The lack of system-level domain knowledge}, which often leads to misinterpretation during log purification.
When directly applying LLMs to log purification, the models tend to produce hallucinations due to their limited understanding of system-specific rules and implicit domain knowledge embedded in logs. Such knowledge is difficult to capture without expert guidance~\cite{huang2025logrules, liu2025r}, resulting in unreliable judgments of sequence purity and anomaly patterns.

To tackle these challenges, we propose \textsc{LogPurge}, a cost-aware and rule-enhanced purification framework that leverages LLMs to purify log sequence training sets, enabling lightweight models trained on the purified subsets to achieve stronger performance.
\textsc{LogPurge} operates in two stages: \textbf{rule-enhanced purification} and \textbf{refined subdivision purification}. In the first stage, \textsc{LogPurge} selects representative sequences within clusters to reduce LLM inference costs, and analyzes them using the LLM to identify and remove high-contamination clusters, while systematically inducing domain rules that enhance the LLM's semantic understanding of logs. In the second stage, motivated by the observation that the remaining low-contamination clusters can be further subdivided to reveal distinct contamination patterns, we decompose the low-contamination purification problem into multiple high-contamination subproblems, each addressed using the strategy of the first stage. Owing to this divide-and-conquer strategy, the problem can be solved without relying on any global or cluster-level anomaly ratio.
Through this hierarchical process, \textsc{LogPurge} progressively eliminates contamination and produces a high-purity dataset, facilitating more effective anomaly detection and model training.
Our contributions are threefold:

\begin{itemize}
    \item We propose a novel LLM-based log purification framework that improves data quality and enables lightweight anomaly detection models to learn diverse log semantics from purified data,
establishing a new paradigm for boosting their effectiveness. 
    \item We design a rule-enhanced LLM evaluator to identify high-contamination regions. By integrating system-level rules into the LLM's prompt and reasoning workflow, we enhance its capability for log purification. These rules, distilled from purification results on the validation set, explicitly encode domain knowledge and introduce targeted semantic constraints, guiding the LLM toward more accurate and consistent judgments.
    \item We introduce a Divide-and-Conquer purification strategy for handling low-contamination regions. By recursively decomposing remaining low-contamination regions into several finer-grained sub-regions and re-evaluating them, our method eliminates residual contamination while preserving valuable semantic diversity.
    \item Our experiments are conducted on two public log datasets and a real-world industry dataset. \textsc{LogPurge} demonstrated exceptional performance, retaining an average of 82.39\% of normal sequences while removing 98.74\% of anomalies. This purification led to significant F-1 score improvements for lightweight anomaly detection models: 35.7\% and 84.11\% on public datasets, and an impressive 149.72\% on the private industrial dataset, thus proving our approach's effectiveness.
\end{itemize}

\section{Background}
\subsection{Preliminaries}
\begin{figure}[t]
  \centering
  \includegraphics[width=\columnwidth]{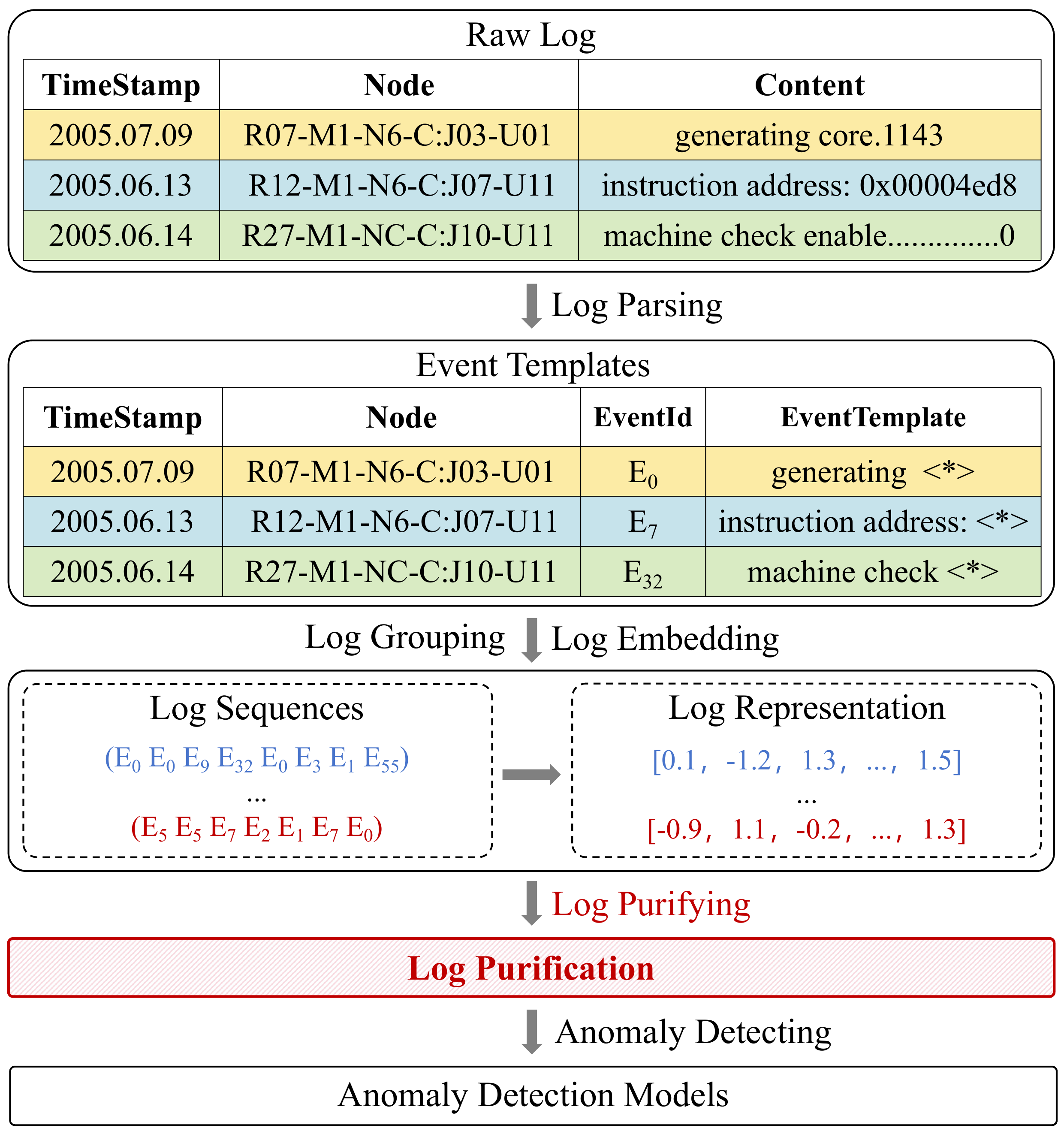}
  \caption{Log Anomaly Detection: The Common Workflow with Log Purification Task}
  \label{fig:workflow}
\end{figure}

In this section, we provide an overview of the key components commonly involved in the general workflow of log analysis tasks. This workflow typically consists of four main steps: (1) log parsing, (2) log grouping, (3) log representation and (4) log anomaly detection, as shown in Figure~\ref{fig:workflow}.

\subsubsection{Log parsing}
Raw logs typically contain elements such as timestamps, component names, log levels, and textual descriptions, which often vary in format and include redundant or noisy information. Directly feeding such unprocessed logs into models may result in sparse features and interference from irrelevant noise. To address this issue, numerous log parsing methods have been proposed to extract parameters and identify event templates, thereby converting raw logs into structured representations~\cite{li2025adaptive,chen2023parser,zhong2024logparser,du2016spell}.
For instance, the log entry
"generating core.1143"
can be parsed into the log template $E_0$: "generating <*>".
Among these approaches, we adopt Drain~\cite{he2017drain} as our log parsing method due to its balance between efficiency and accuracy.

\subsubsection{Log grouping}
The log grouping strategy can be adapted to data characteristics. Common approaches include: (1) Fixed window: logs are divided by a predefined time span or number of entries; (2) Sliding window: the window moves along the log sequence with a fixed step, producing overlapping groups; (3) Session window: logs are segmented by unique identifiers. In this study, we adopt a sliding-window strategy to construct log sequences consistently across datasets.

\subsubsection{Log representation}
After log grouping, it is necessary to transform logs into representations compatible with the input requirements of anomaly detection models. In this study, we utilize the pre-trained language model BERT to convert each log sequence into a fixed-dimensional vector, thereby capturing the semantic information contained in the logs. The resulting vectorized representations preserve both the semantic meaning and the event pattern characteristics of the log sequences.

\subsubsection{Log anomaly detection}
After converting raw log sequences into vector embeddings, these are fed into anomaly detection models. Traditional unsupervised methods~\cite{xu2009largescale, liu2008isolation, breunig2000lof} learn normal patterns to detect deviations, while recent deep learning approaches~\cite{du2017deeplog, yang2021plelog, guo2021logbert, li2022unsupervised, wang2021multi, zhang2022cat} capture richer temporal and semantic dependencies for more accurate and robust detection.
\subsection{Motivation}
\subsubsection{Limitations for \textsc{Pluto}}
\label{sec:limits-pluto}
\textsc{Pluto} is, to the best of our knowledge, the only method that explicitly tackles log purification in an unsupervised setting. Pluto partitions the embedding space into several regions, where both high- and low- contamination regions are observed(as shown in Figure ~\ref{fig:tsne_distribution}). For each cluster, given its embedding matrix \(E\), it computes the first and second singular values \(\lambda_1\) and \(\lambda_2\) (\(\lambda_1 > \lambda_2\)) to capture structural characteristics. The \emph{dominance} is defined as
\[
  \mathrm{dom} = \frac{\lambda_1}{\lambda_2}.
\]

Based on the dominance value, \textsc{Pluto} distinguishes high contamination and low contamination regions in the embedding space by finding a spike in the dominance curve. In particular, clusters with high dominance are regarded as highly contaminated, while those with low dominance are treated as relatively clean. However, a low dominance value may also arise from opposite patterns within the same cluster, which indicates that the corresponding subspace aligns with normal behavior.
Once the low-contamination regions are identified, \textsc{Pluto} further refines them by removing samples that align with the second singular vector and are equal in number to the estimated contaminated samples within the cluster. To perform this step, it first estimates the anomaly ratio for each cluster, where the estimation follows opposite rules for high- and low-contamination groups:
\[
r_i =
\begin{cases}
 \dfrac{1}{\mathrm{dom}_i} \cdot \alpha, & \text{if cluster is low-contamination}, \\
 \mathrm{dom}_i \cdot \alpha, & \text{if cluster is high-contamination},
\end{cases}
\]
where \(\alpha\) is a scaling coefficient. 
Since the formulas are opposite, misclassifications in the first step directly invalidate the estimation. 
\begin{figure}[t]
  \centering
  \includegraphics[width=\columnwidth]{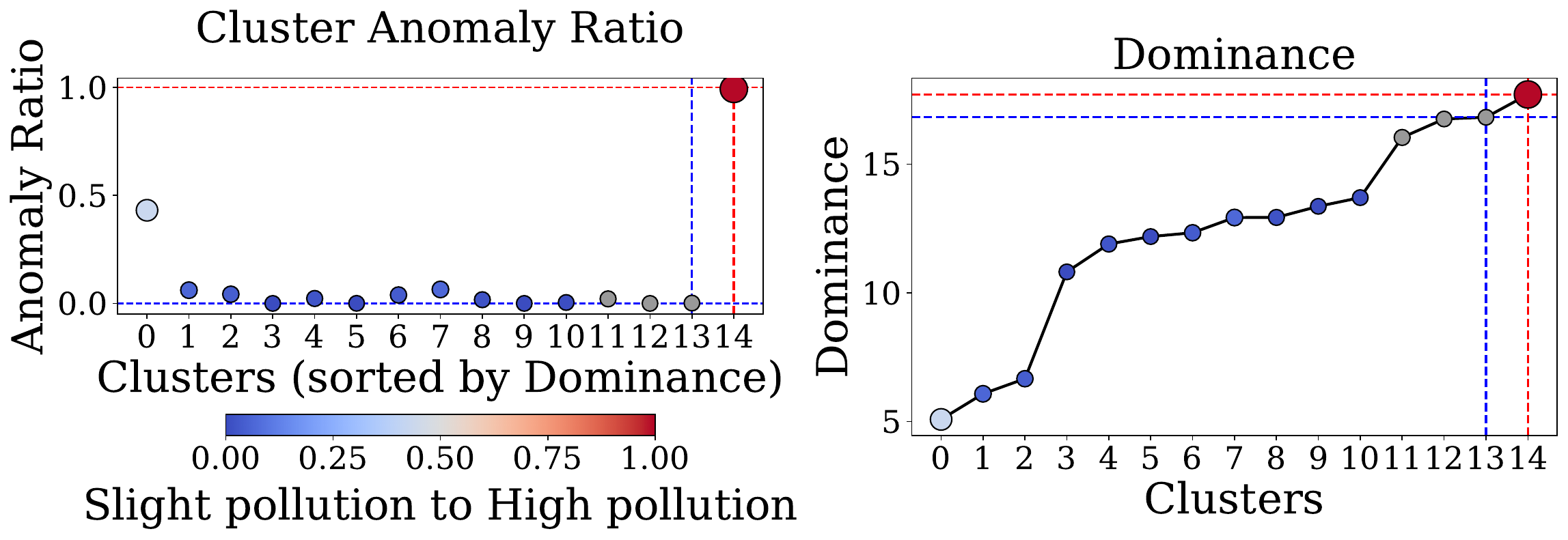}
  \caption{Cluster anomaly ratio and dominance on the Liberty log sequence dataset. 
  We adopt the same embedding procedure and dominance computation as used in \textsc{Pluto}. 
  Since the original \textsc{Pluto} implementation does not provide a data preprocessing pipeline, 
  we follow the preprocessing code from its underlying base model, \textsc{LogBERT}, to prepare the input sequences.}
  \label{fig:domin}
\end{figure}
Empirical results show that the assumption does not always hold (see Figure~\ref{fig:domin}). 
In cluster~14, although it has the highest dominance value, this cluster actually mixes both normal and abnormal templates that frequently co-occur. 
Specifically, the anomalous event  
\texttt{pbs\_mom: Bad file descriptor (9) in <*><*><*>}  
appears in 99.80\% of sequences, while the normal template  
\texttt{pbs\_mom: pbs\_mom, wait\_request failed}  
occurs even more frequently (99.82\%).  
Such coexistence leads to an extremely high anomaly ratio, yet its dominance value is not substantially higher than that of other clusters. 
In contrast, clusters~11, 12, and 13 contain almost no anomalies, but they still exhibit high dominance because their embeddings are concentrated around a few frequently occurring normal templates. 
These observations suggest that high dominance can result from either abnormal or normal template concentration. 
Therefore, dominance alone is insufficient to distinguish high- and low-contamination clusters, and incorrectly filtering low-contamination clusters under this criterion can significantly degrade unsupervised anomaly detection performance.

Once the high- and low-contamination clusters are incorrectly identified, the subsequent estimation of anomaly ratios for each cluster will become completely biased in the opposite direction.
This misestimation leads the low-contamination strategy to mistakenly remove samples that are aligned with the secondary embedding vector, ultimately causing the purification process to fail.
Moreover, accurate estimation of each cluster's anomaly ratio relies on a precise global anomaly ratio, which is nearly impossible to obtain in real-world applications.

\begin{samepage}
\begin{mdframed}[backgroundcolor=findingsgray,linewidth=0.5pt,roundcorner=3pt, skipabove=8pt, skipbelow=8pt]
\textbf{Findings:} Dominance measures \emph{pattern concentration}, not \emph{pattern semantics}. 
Without semantic evidence, Pluto cannot determine whether the dominant component is abnormal or normal, leading to erroneous estimates. 
Therefore, to achieve robust purification, it is necessary to perform \emph{semantic analysis of clusters}.
\end{mdframed}
\end{samepage}

\subsubsection{Solution by Using Rule-based LLM}
\label{sec:Solution by Using Rule-based LLM}
Inspired by the findings in Section~\ref{sec:limits-pluto}, we conduct a small-scale experiment to address Pluto's misclassifications using an LLM. Specifically, we select the cluster that Pluto has incorrectly labeled and ask GPT-4o to determine whether the cluster represented a high- or low-contamination region.

\begin{figure}[t]
  \centering
  \includegraphics[width=1.05\columnwidth]{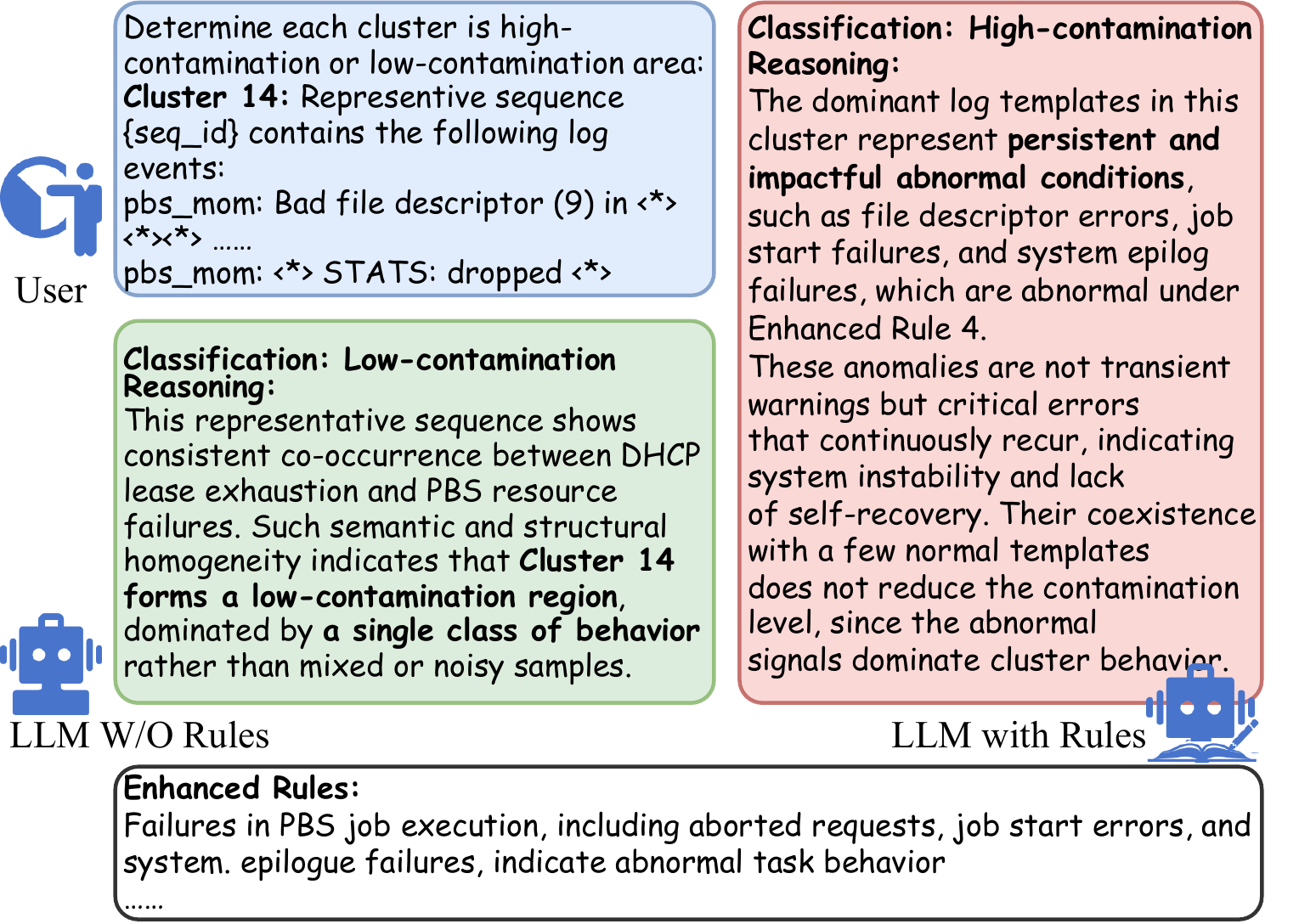}
  \caption{Example of LLM reasoning for judging high-contamination areas.}
  \label{fig:llm-analysis}
\end{figure}
Without additional guidance, the LLM initially exhibited inconsistent reasoning. For example, it classified the cluster with clear "Bad file descriptor" patterns as a low-contamination region, since it focused mainly on surface-level co-occurrences rather than the severity of abnormal behaviors. Nevertheless, it still demonstrated a strong grasp of log semantics-capturing structural relationships among log events and identifying general anomaly patterns with interpretability (Figure~\ref{fig:llm-analysis}), which meets engineers' needs~\cite{zhao2021empirical}. After integrating rule-based enhancements distilled from system domain knowledge (e.g., "persistent PBS execution failures imply abnormal task behavior"), the LLM reasoning became more systematic and accurate. It correctly recognized that the cluster dominated by recurring system-level failures corresponded to a high-contamination region, providing explicit causal explanations for its decisions.

This experiment highlights a fundamental limitation of Pluto: its statistical nature prevents it from explaining why certain clusters exhibit abnormality. In contrast, LLMs-especially when augmented with explicit rules-combine statistical and semantic perspectives, bridging quantitative anomaly ratios with interpretable domain reasoning. These insights motivate our framework design, which leverages rule-enhanced LLM evaluation to refine contamination judgment and improve the reliability of downstream anomaly detection.

\begin{mdframed}[backgroundcolor=findingsgray,linewidth=0.5pt,roundcorner=3pt,  skipabove=8pt,
  skipbelow=8pt]
\textbf{Findings:} Incorporating rules enables LLMs to leverage domain knowledge effectively for log purification, significantly improving classification accuracy. 
\end{mdframed}

\subsubsection{Motivation for LogPurge}

Unsupervised log anomaly detection requires both sufficient normal logs and the removal of contaminated anomalies, yet meeting these conditions in practice is extremely challenging. To overcome this, we exploit the semantic understanding of LLMs to assist small classifiers in improving data quality. By providing rule-guided semantic judgments, LLMs help the small model retain more valid normal logs while filtering out harmful anomalies. This collaboration enhances the efficiency and applicability of unsupervised log anomaly detection, maximizing the complementary strengths of LLMs and small models.
\section{Approach}
\begin{figure*}[t]
  \centering
  \includegraphics[width=1.0\linewidth]{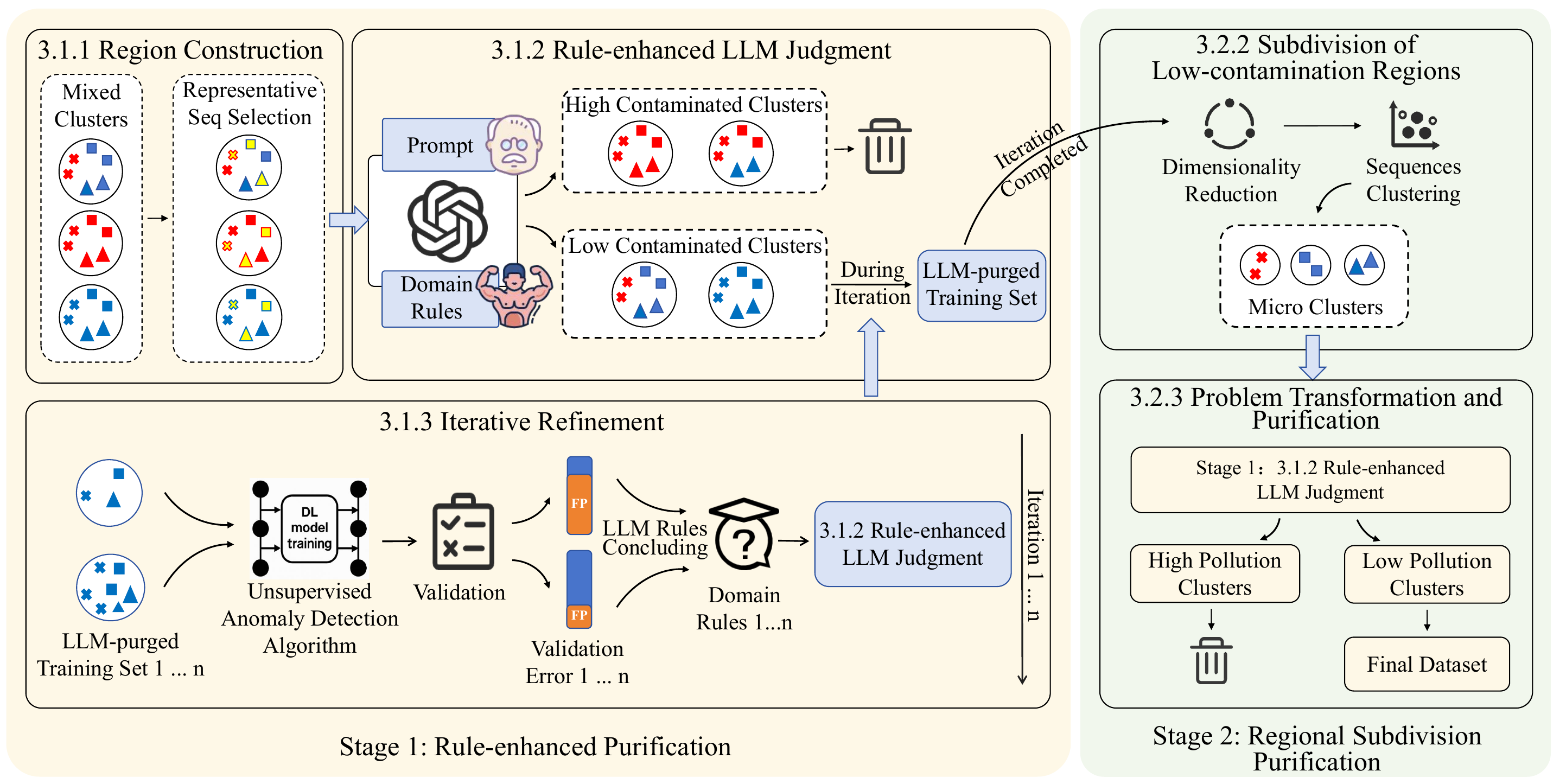}
  \caption{The overall architecture of LogPurge}
  \label{fig:architecture}
\end{figure*}

\noindent
To address the reliance on labels in unsupervised log anomaly detection and the challenges of high inference cost and limited domain knowledge in LLM-based log analysis, we propose LogPurge, a hierarchical iterative purification framework. Figure ~\ref{fig:architecture} shows the architecture of LogPurge. LogPurge consists of two stages: \textit{rule-enhanced purification} and \textit{refined subdivision purification}. In the first stage, we select representative sequences are clustered and analyzed by the LLM to remove high-contamination clusters and induce systematic domain rules; in the second stage, the remaining low-contamination clusters are further subdivided and recursively purified. Through this hierarchical process, LogPurge progressively eliminates contamination and yields a high purity dataset for more effective anomaly detection and model training.

\subsection{Rule-enhanced Purification}
\label{sec:Rule-enhanced Purification}
To effectively eliminate high-contamination regions, we design an iterative purification strategy that progressively improves the understanding of domain-specific knowledge of the LLM, enabling accurate removal of polluted clusters. The rule-enhanced purification stage consists of three steps: region construction, rule-enhanced LLM judgment, and iterative refinement.
\subsubsection{Region Construction}
\label{sec:Region Construction}
Given a set of log sequence embeddings 
\(\mathcal{X} = \{\mathbf{x}_1, \mathbf{x}_2, \dots, \mathbf{x}_N\}\), 
where each \(\mathbf{x}_i \in \mathbb{R}^d\) denotes the embedding vector of a log sequence obtained from the pretrained BERT~\cite{kenton2019bert}, 
we first partition the embedding space into regions 
\(\{\mathcal{R}_1, \mathcal{R}_2, \dots, \mathcal{R}_K\}\) 
using K-means clustering. 
To ensure semantic coverage while avoiding redundancy, 
we then select representative sequences based on a density-peak principle.

For each sample \(\mathbf{x}_i \in \mathcal{R}_k\), its local density is defined as
\[
\rho_i = \frac{1}{\overline{d}_i + \epsilon},
\]
where \(\overline{d}_i\) denotes the average distance to its \(k\) nearest neighbors 
and \(\epsilon\) is a smoothing factor. 
A point is chosen as a regional center \(\mathbf{c}_k\) if it exhibits high density 
and is separated from existing centers by at least \(r_{\min}\). 
Around each center, a small number of neighbors are further selected 
to form a representative set \(\mathcal{S}_k\).

The final representative set
\[
\mathcal{S} = \bigcup_{k=1}^{K} \mathcal{S}_k
\]
captures the essential characteristics of all regions, enabling subsequent LLM reasoning
to operate only on \(|\mathcal{S}| \ll N\) samples, thus preserving semantic fidelity
while substantially reducing inference cost.
\subsubsection{Rule-enhanced LLM Judgment}
After selecting the representative sequences, we employ LLMs to assess the level of contamination in each region. The representative set is defined as
$\mathcal{S}$
where 
\(\mathcal{S}_k = \{\mathbf{s}_{k1}, \mathbf{s}_{k2}, \dots, \mathbf{s}_{kM_k}\}\) 
denotes the representative sequences of region \(\mathcal{R}_k\). For each representative sequence \(\mathbf{s}_{kj}\), we construct an input prompt \(\pi(\mathbf{s}_{kj})\) and query the LLM to obtain a predicted label, as our experiments in Section ~\ref{sec:Solution by Using Rule-based LLM}:
\[
\hat{y}_{kj} = f_{\text{LLM}}(\pi(\mathbf{s}_{kj})),
\]
where \(\hat{y}_{kj} \in \{\text{low-contamination}, \text{high-contamination}\}\).
Figure~\ref{fig:llm-analysis} illustrates an example of the prompt format we employ.

Once each region's predicted label is obtained, if region \(\mathcal{R}_k\) is classified as low-contamination, i.e., \(\hat{y}_k = \text{low-contamination}\), all sequences in that region are added to the training set:
\[
\mathcal{T} \leftarrow \mathcal{T} \cup \mathcal{R}_k.
\]

The constructed training set $\mathcal{T}$ is then used to train a lightweight anomaly detection model, which serves as a cost-efficient detector. To evaluate the data quality ability of the purified data, we perform validation on a separate validation set $\mathcal{V}$ using this lightweight model trained on $\mathcal{T}$. For any sample $\mathbf{v} \in \mathcal{V}$ whose predicted label differs from the ground-truth:
\[
\hat{y} \neq y,
\]
the sample is added to an error set $\mathcal{E}$. 
Previous results~\cite{cui2025logeval,liu2024logprompt,qi2023loggpt} and our experiments(Section~\ref{RQ3:Cost-Effectiveness of LLM-based Purification for Anomaly Detection}) show that LLMs tend to overreact to semantic deviations, producing false positives for logs that do not represent real anomalies. 
Interestingly, this sensitivity inspires unsupervised validation: since unsupervised models typically rely on only a few normal samples, such LLM-induced false positives provide useful boundary cases for refining validation rules and thresholds. 
Following the representative selection principle in region construction, we select a subset $\mathcal{E}^* \subseteq \mathcal{E}$ and feed it back to the LLM to to summarize system-level domain rules:
\[
\mathcal{R}_{\text{domain}} = f_{\text{LLM}}(\mathcal{E}^*).
\]

As a result, low-contamination regions contribute additional data to the training set, while validation errors are transformed into domain rules, completing one cycle of LLM-based contamination assessment.

\subsubsection{Iterative Refinement}

To further improve the reliability of the contamination assessment, we adopt an iterative refinement strategy. In each iteration, the current domain rule set
\(\mathcal{R}_{\text{domain}}\) is incorporated into the region partitioning and LLM evaluation process. As the iterations proceed, new misclassified samples are identified and summarized into additional rules, leading to an incremental expansion of the rule set:
\[
\mathcal{R}_{\text{domain}}^{(t+1)} = \mathcal{R}_{\text{domain}}^{(t)} \cup f_{\text{LLM}}(\mathcal{E}^{*(t)}),
\]
where \(\mathcal{E}^{*(t)}\) denotes the representative subset of misclassified validation samples in the \(t\)-th iteration.

With the accumulation of domain rules, the LLM's judgments become progressively more stable. We define convergence when the set of low-contamination regions no longer expands, i.e.,
\[
\mathcal{R}_{\text{low}}^{(t+1)} = \mathcal{R}_{\text{low}}^{(t)},
\]
which is equivalent to saying that the predicted region labels show no significant changes between consecutive iterations. Once this condition is met, the iteration terminates. At this point, high contamination regions have been effectively removed, and the remaining dataset can serve as high-quality training data for downstream anomaly detection models.
\subsection{Regional Subdivision Purification}

To eliminate residual contamination in low-contamination regions, we design a Divide-and-Conquer purification strategy
that leverages the domain-specific knowledge already accumulated by the LLM, enabling direct and accurate judgment
without additional iterations. The regional subdivision purification stage consists of two steps:
first, subdividing low-contamination regions into finer-grained sub-regions to better isolate potential anomalies; and second, transforming the purification task into multiple high-contamination subproblems, just as the challenge we addressed in Section \ref{sec:Rule-enhanced Purification}.

\subsubsection{Observation of Residual Clustering}
\label{sec:Observation of Residual Clustering}
\begin{figure}[htbp]
    \centering
    \includegraphics[width=0.88\linewidth]{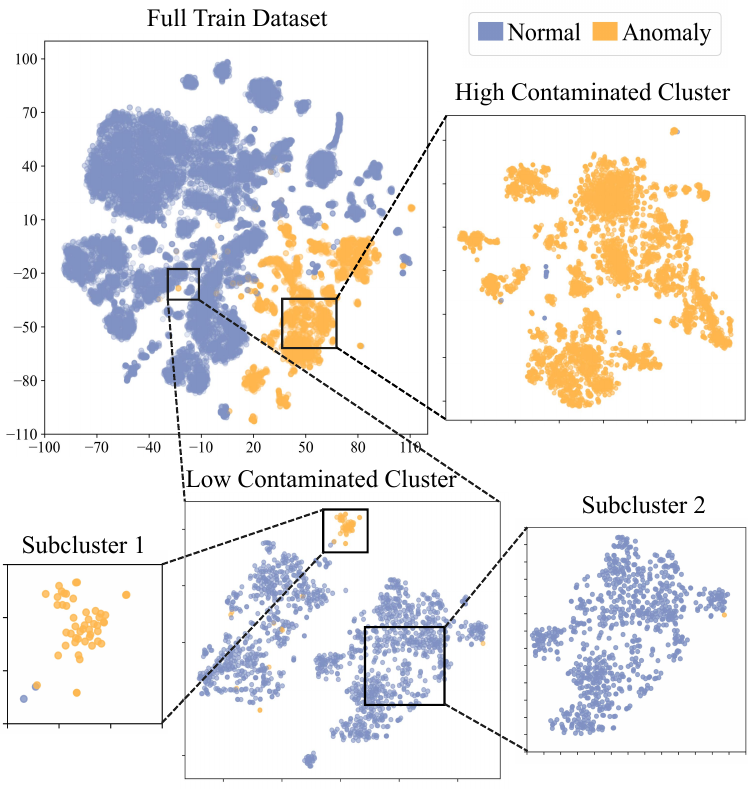}
    \caption{t-SNE visualization of Liberty training set, highlighting the clustering of anomalous sequences and their distribution across low and high contamination regions.}
    \label{fig:tsne_distribution}
\end{figure}
After performing t-SNE~\cite{maaten2008visualizing} dimensionality reduction on the training set Liberty as shown in Figure ~\ref{fig:tsne_distribution}, the resulting visualization clearly shows the overall distribution of the data, particularly the distribution patterns of anomalous sequences. The plot reveals a significant clustering of anomalous sequences (as shown in the top-right subplot). This clustering phenomenon aligns with the findings in Pluto's analysis, suggesting that when the system experiences anomalies, a large number of anomaly alerts are generated, causing the anomalous logs to be concentrated in specific regions. It's important to note that these clusters do not imply that all anomalous sequences share the same anomaly pattern; in fact, due to the system's heterogeneity, these clusters do not always follow a single anomaly pattern.

In the low-contamination regions, we conducted a further dimensionality reduction and discovered that anomalous log sequences still exhibit similar high and low contamination distributions, akin to the overall dataset. This phenomenon can be explained by the fact that anomalous log sequences often contain only a small number of anomalies. These sequences, while similar to normal logs in the overall distribution, still show a slight divergence due to the presence of a few anomalous entries. In finer sub-region divisions, these small amounts of anomalous logs do not align well with the principal components, causing them to form separate clusters that are distinct from the normal sequence distribution. This suggests a Divide and Conquer approach: breaking the problem into sub-region subproblems.

\subsubsection{Subdivision of Low-contamination Regions}

Guided by the observations discussed above, we perform region construction based on the t-SNE visualization results of the BGL training set. Specifically, we apply t-SNE to project high-dimensional log representations into a two-dimensional latent space, where structurally similar log sequences are placed closer together while dissimilar ones are pushed apart. This projection preserves the local neighborhood relationships of the original high-dimensional space, making it particularly suitable for identifying clusters or sub-regions that correspond to distinct behavioral patterns in the system.

Logs are inherently high-dimensional and sparse due to diverse event templates and variable contextual dependencies. Directly analyzing such data in the original feature space often leads to fragmented or misleading clustering results. By contrast, t-SNE effectively captures the manifold structure underlying the log representations, allowing visually coherent regions to emerge in the low-dimensional space. These regions intuitively reflect groups of logs that share similar semantics or operational states.

Based on this reduced representation, we partition each low-contamination region $\mathcal{R}_{\text{low}}$ into finer-grained sub-regions:
\[
\mathcal{R}_{\text{low}} \;\longrightarrow\; \{\mathcal{R}_{\text{low},1}, \mathcal{R}_{\text{low},2}, \dots \}.
\]
This subdivision enables residual anomalies-previously diluted within a broad region-to be concentrated within smaller sub-regions, making them more distinguishable from normal behaviors. In this way, the t-SNE-based regional construction provides an observation-driven and semantically consistent framework for capturing subtle anomaly patterns that may otherwise be overlooked in the original space.

\subsubsection{Problem Transformation and Purification}
Once subdivided, each subregion is directly evaluated by the LLM using the accumulated domain rules.
High-contamination sub-regions are eliminated, while clean sub-regions are retained and added to the training set.
Through this process, the residual contamination problem in low-contamination regions is effectively
transformed back into the same form addressed in the rule-enhanced purification stage,
allowing systematic removal without additional iterations.

\section{Experiment}
\subsection{Research Questions}

\begin{itemize}
    \item \textbf{RQ1:} How effective is the LLM-based purification framework in removing contaminated samples from log training datasets?
    \item \textbf{RQ2:} To what extent does purified data improve the performance of lightweight anomaly detection models?
    \item \textbf{RQ3:} Compared to directly applying LLMs for anomaly detection, is the LLM-based purification more advantageous in terms of cost and effectiveness?
    \item \textbf{RQ4:} What are the contributions of the main components in the purification framework?
\end{itemize}

\subsection{Experiment Setup}
\subsubsection{Datasets}
 We conduct experiments using two public log datasets: BGL, Liberty, and a software system log datasets from the
production environments of a top-tier global ISP collected in
2023
which we selected following the related works~\cite{oliner2007supercomputers,zhu2023loghub,hashemi2024onelog,zhang2024end}. Detailed information about these datasets is provided in
Table~\ref{table: details of datasets}. For Liberty, we randomly sampled 5 million log entries as our dataset. For all datasets, we adopted the mainstream grouping approach~\cite{guo2021logbert,yang2021plelog}, using a sliding window with a length of 60 seconds and a stride of 30 seconds to divide logs into sequences. These sequences were then split into training, validation, and test sets in a 6:2:2 ratio. During evaluation, both LogPurge and baseline methods selected identical log sequences as training samples for model input.

\begin{table}[t]
\centering
\caption{Detailed information of the datasets}
\label{table: details of datasets}
\begin{tabular}{lcccc}
\toprule
\textbf{Dataset} & \textbf{Count} & \textbf{Anomalies} & \textbf{Train Seq} & \makecell{\textbf{Abnormal}\\\textbf{Train Seq}}\\
\midrule
BGL & 4,747,963 & 348,460 & 43,481 & 3,140\\
Liberty & 5,000,000 & 1,814,386 & 26,102 & 4,939\\
Industry & 1,187,117 & 249,955 & 14,014 & 7,563\\
\bottomrule
\end{tabular}
\end{table}

\subsubsection{Baselines}

We compare the performance of LogPurge with three different sample selection methods followed by anomaly detection, as well as with two traditional machine learning methods.

\textbf{Machine Learning Methods.}
Traditional unsupervised machine learning methods remain representative in log anomaly detection, as they typically rely on statistical characteristics or feature space partitioning to identify anomalies. In our experiments, we compare two widely used methods: PCA ~\cite{xu2009largescale} and IsolationForest ~\cite{liu2008isolation}. PCA performs dimensionality reduction through principal component analysis of log features and uses reconstruction error as an anomaly measure. In contrast, IsolationForest incrementally partitions the feature space at random, where anomalous points usually require fewer splits to be isolated, making them effectively detectable. Both are directly applied to the original, unpurified training set to serve as baselines for evaluating performance gains from dataset cleaning.

\textbf{Sample Selection Methods.}
As our proposed method, LogPurge, is an LLM-based training set purification approach, we compare it with three representative sample selection methods designed to mitigate training set contamination: FINE~\cite{kim2021fine}, Pluto~\cite{ma2024pluto}, and LogCleaner~\cite{zhang2024reducing}. FINE constructs a class-wise Gram matrix to extract principal components via eigen decomposition, and identifies well-aligned samples using FINE scores and a Gaussian Mixture Model. Pluto partitions the embedding space into high- and low-contamination regions and filters anomalous clusters or samples based on dominance and feature alignment metrics. LogCleaner leverages TF-IDF, mutual information, and OPTICS clustering to remove anti- and duplicative-events, producing a concise dataset adaptable to system changes.

\textbf{Lightweight Anomaly Detection Methods.}
To evaluate the effectiveness of training set purification, we adopt three representative deep learning-based log anomaly detection methods: DeepLog~\cite{du2017deeplog}, LogBERT~\cite{guo2021logbert}, and PLELog~\cite{yang2021plelog}. DeepLog models log sequences using an LSTM architecture and detects anomalies by predicting the next log event. LogBERT leverages the Transformer architecture with a masked prediction task to learn log representations, effectively capturing global dependencies. PLELog processes unstable log data through semantic embeddings and employs an attention-based GRU network for anomaly detection.

\textbf{LLM-based Anomaly Detection Methods.}
To evaluate the cost-effectiveness and performance of our proposed method, we compare it against the original LLM model, GPT-4o~\cite{achiam2023gpt}. Additionally, we include two LLM-based log anomaly detection methods that have influenced our approach: LogPrompt~\cite{liu2024logprompt} and LogRules~\cite{huang2025logrules}. LogPrompt leverages a specialized workflow for log template detection, while LogRules demonstrates the effectiveness of using specific log rules to enhance the performance of log anomaly detection through GPT-4o.

\textbf{Implementation Details of Baseline Methods.}
We briefly describe the implementation details of the three sample selection methods. Both FINE and Pluto adopt K-means clustering with the number of clusters set to 20, and the anomaly rate fixed to the actual anomaly rate of the dataset. For LogCleaner, the TF-IDF threshold for filtering low-frequency events is set to 0.1, the mutual information threshold $\theta_{anti}$ to 0, and the OPTICS clustering parameter $\theta_{dup}$ to 5. With these parameter configurations, the three methods are applied to purify the dataset and reduce contamination.
After sample selection, we evaluate the effectiveness using three log anomaly detection models: DeepLog, LogBERT, and PLELog. DeepLog and LogBERT are implemented following LogDeep~\cite{le2022log}. Specifically, DeepLog employs a two-layer LSTM with a hidden dimension of 64, while LogBERT uses a two-layer Transformer with a hidden size of 256 and 4 attention heads. PLELog adopts a three-layer GRU network with a hidden dimension of 64. In Section~\ref{RQ3:Cost-Effectiveness of LLM-based Purification for Anomaly Detection}, we evaluate our method using LLM-based log anomaly detection approaches, with the temperature set to 0 to ensure consistent responses.

\subsubsection{Evaluation metrics}
To quantitatively evaluate the performance of our clean data selection framework, 
we design two categories of evaluation metrics: (1) clean subset selection metrics and (2) anomaly detection metrics.
These metrics are theoretically grounded in sample selection methods~\cite{song2019selfie,wu2020topological,ma2024pluto}.

\paragraph{(1) Clean Subset Selection Metrics.}
We first evaluate the purity and completeness of the selected clean subset.

\textbf{Subset Purity (SP)} measures the proportion of \textbf{normal (clean) samples} retained in the selected subset:
\begin{equation}
\text{SP} = \frac{\left|S_{\text{selected}} \cap S_{\text{normal}}\right|}{\left|S_{\text{selected}}\right|}
\end{equation}
A higher SP indicates higher subset purity and stronger robustness against contaminated data.

\textbf{Clean Retention Rate (CRR)} quantifies how many truly normal samples are preserved:
\begin{equation}
\text{CRR} = \frac{|S_{\text{selected}} \cap S_{\text{clean}}|}{|S_{\text{clean}}|}
\end{equation}
A higher CRR implies better preservation of valuable clean data for downstream model training.

SP and CRR together characterize the trade-off between data purity and completeness, related closely to the two goals of log purification task.

\paragraph{(2) Anomaly Detection Metrics.}
Following standard practice in evaluating purified datasets for downstream anomaly detection, 
we report \textit{Precision}, \textit{Recall}, and \textit{F1-score} 
to measure how effectively the purified data supports accurate anomaly identification.

Let $TP$ denote the number of correctly detected abnormal samples, 
$FP$ denote the normal samples incorrectly identified as abnormal, 
and $FN$ denote the abnormal samples that remain undetected. 
The three metrics are defined as follows:
\begin{equation}
\text{Precision} = \frac{TP}{TP + FP}
\end{equation}
\begin{equation}
\text{Recall} = \frac{TP}{TP + FN}
\end{equation}
\begin{equation}
\text{F1-Score} = \frac{2 \times \text{Precision} \times \text{Recall}}{\text{Precision} + \text{Recall}}
\end{equation}

Here, \textit{Precision} reflects the purity of detected anomalies, while \textit{Recall} indicates the completeness of anomaly discovery. 
In log anomaly detection tasks, the class distribution is often highly imbalanced (i.e., anomalies are rare). 
Under such imbalance, the \textit{F1-score}, as the harmonic mean of precision and recall, provides a more reliable single-value summary of detection performance because it jointly penalizes low precision or low recall and thus better reflects the trade-off between false alarms and missed detections.

\subsection{RQ1:Effectiveness in Removing Contamination}\label{RQ1:Effectiveness in Removing Contamination}

The experimental results aim to evaluate the efficacy of various log cleaning methods in terms of dataset Purity (SP) and Correct Retention Rate (CRR). These two metrics are essential for assessing the performance of log cleaning techniques, as they directly impact the quality of data used for downstream anomaly detection tasks. In particular, while high SP indicates a cleaner dataset by removing anomalous logs, it is equally important to preserve as many normal logs as possible, which is reflected by a higher CRR. A method that achieves high SP but sacrifices too many normal logs will lead to poor anomaly detection performance due to the lack of representative normal data for training.

As shown in Table~\ref{tab:eq1}, the original data, without any preprocessing, indicates significant imbalance between normal and anomalous logs. However, CRR is not applicable here, as no log filtering or selection has been performed.

FINE, a traditional sample selection method, performs well in terms of SP, reaching 96.47\% on BGL, but fails to retain a sufficient number of normal logs, resulting in a CRR of 25.56\% on BGL, which means it only keeps about 30\% normal data compared to LogPurge. This is due to FINE's reliance on principal component alignment, which is less effective in high contamination areas. In these regions, anomalous logs dominate the principal components, causing FINE to discard a significant portion of normal logs. This observation highlights the limitation of FINE in datasets with high contamination, where its sample selection strategy becomes ineffective.

On the other hand, Pluto achieves the highest SP values (e.g., 98.72\% on BGL and 93.98\% on Industry) among the baseline methods. However, its CRR remains low (e.g., 40.16\% on BGL), reflecting an overly aggressive filtering mechanism. By focusing solely on anomalous cluster removal through statistical dominance, Pluto sacrifices a large number of normal logs, resulting in a low CRR and potentially detrimental effects for anomaly detection models.

In contrast, LogCleaner, an event-level method, performs poorly due to its tendency to remove entire sequences when an event is misidentified. As a result, LogCleaner achieves the lowest SP and CRR among the methods, with values as low as 66.46\% for SP and 21.62\% for CRR on the Industry dataset. This emphasizes the severe limitations of event-level processing, which overlooks critical sequence-level dependencies.

\textbf{LogPurge} combines both statistical distribution analysis and semantic understanding of log sequences. This combination allows LogPurge to achieve high SP values (e.g., 99.58\% on Liberty) while maintaining excellent CRR (e.g., 92.35\% on Liberty). The balance between high purity and high retention of normal logs makes LogPurge the most effective method, ensuring both a clean dataset and sufficient normal data for training anomaly detection models.

In conclusion, SP and CRR must be considered together when evaluating log cleaning methods. While high SP indicates a cleaner dataset, it is the retention of normal logs, as reflected by CRR, that ensures the effectiveness of downstream anomaly detection tasks. LogPurge provides the best balance, achieving superior performance across both metrics and demonstrating its capability for preparing datasets for anomaly detection applications.

\begin{table}[t!]
    \caption{Clean Subset Selection Metrics of baselines}
    \centering
    \resizebox{\linewidth}{!} {%
    \large
    \begin{tabular}{l@{\hskip 0.1in}c@{\hskip 0.1in}c@{\hskip 0.1in}c@{\hskip 0.1in}c@{\hskip 0.1in}c@{\hskip 0.1in}c@{\hskip 0.1in}c}
    \toprule
    \multirow{2}{*}{\textbf{Method}} & \multicolumn{2}{c}{\hspace{-1.5em}\textbf{BGL}}       & \multicolumn{2}{c}{\hspace{-0.2em}\textbf{Liberty}}  & \multicolumn{2}{c}{\hspace{-0.2em}\textbf{Industry}} 
    \\ \cmidrule(l{0em}r{1em}){2-3} \cmidrule(l{0.7em}r{1.3em}){4-5}
    \cmidrule(l{0em}r{0.7em}){6-7}
                                   & \hspace{-0.1em}\textbf{SP(\%)} & \textbf{CRR(\%)} & \textbf{SP(\%)}   & \textbf{CRR(\%)} & \textbf{SP(\%)} & \textbf{CRR(\%)} \\ \midrule
    Original & 86.01 & - & 81.08 & - & 84.27 & - \\
    \hdashline
    \noalign{\vskip 2pt}
    FINE~\cite{kim2021fine}& 96.47 & 25.56 & 82.38 & 73.23 & 83.82 & 70.94 \\
    Pluto~\cite{ma2024pluto}        & \textbf{98.72} & 40.16 & 92.11 & 41.68 & 93.98 & 48.27\\
    LogCleaner~\cite{zhang2024reducing}            & 93.34 & 45.65 & 87.67 & 57.22 & 66.46 & 21.62 \\
    \hdashline
    \noalign{\vskip 2pt}
    \textbf{LogPurge}        & 97.18 & \textbf{83.14} & \textbf{99.58} & \textbf{92.35} & \textbf{99.47} & \textbf{71.67} \\ 
    \bottomrule 
    \end{tabular}
    }
    \label{tab:eq1}
\end{table}

\begin{figure}[htbp]
    \centering
    \includegraphics[width=\linewidth]{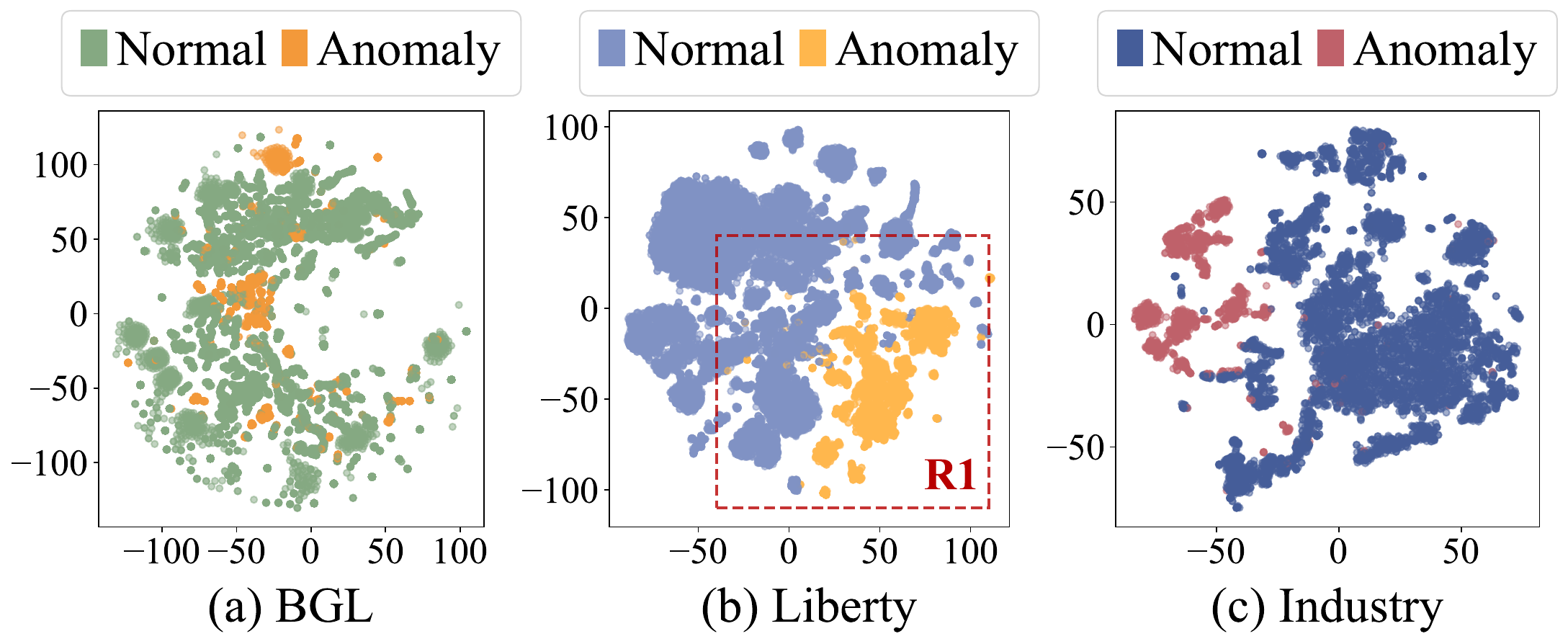}
    \caption{t-SNE visualization of sequence embeddings of the three training sets. R1: a local subset containing both normal and abnormal samples.}
    \label{fig:tsne_distribution_for3_domins}
\end{figure}

\textbf{Dataset Visualization:}
Figures~\ref{fig:tsne_distribution_for3_domins}(a)-(c) show the t-SNE visualization of the sequence embeddings of the three contaminated training sets before log purification. Distinct clustered distribution characteristics can be observed across all datasets-that is, normal and anomaly sequences are not randomly scattered but rather form high-density clusters in the embedding space. This observation is consistent with the findings discussed in Section \ref{sec:limits-pluto} regarding the Pluto method, as well as with the distribution patterns shown in Figure ~\ref{fig:tsne_distribution} of Section \ref{sec:Observation of Residual Clustering}.

\begin{figure}[htbp]
    \centering
    \includegraphics[width=\linewidth]{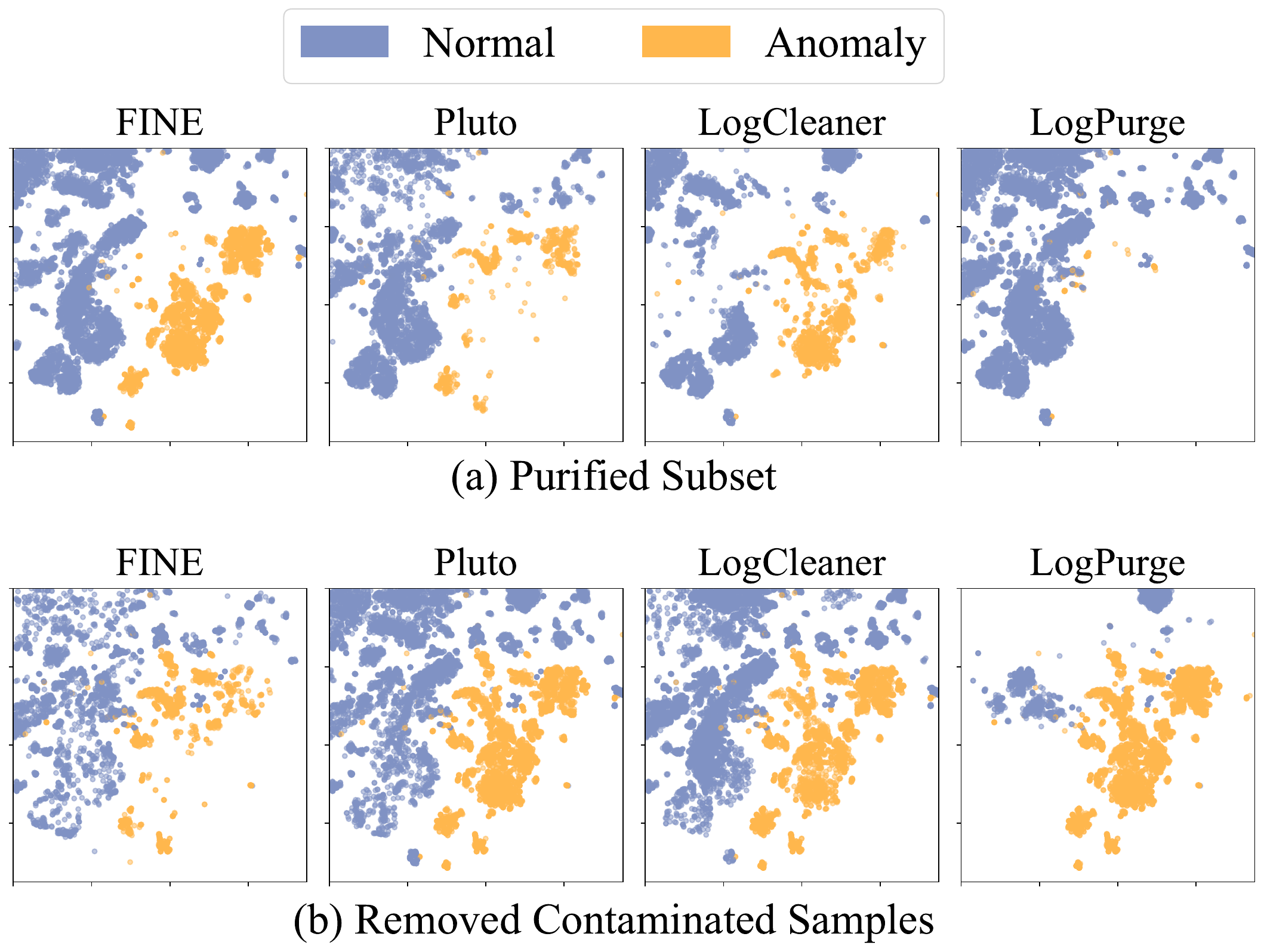}
    \caption{t-SNE visualization of R1 purification effects achieved by four different methods}
    \label{fig:tsne_distribution_4methods}
\end{figure}

\textbf{Purification of Contaminated Samples:} 
From the Liberty dataset shown in Figure~\ref{fig:tsne_distribution_for3_domins}(b), we selected a representative subset region R1, where anomalies are relatively concentrated, to illustrate the effectiveness of different log purification methods, including FINE, Pluto, and LogCleaner.
Figure~\ref{fig:tsne_distribution_4methods}(a) presents the purified results of the four purification methods within the R1 region, while Figure~\ref{fig:tsne_distribution_4methods}(b) shows the contaminated samples removed by these four methods in the same region.
It can be observed that all baseline methods fail both to eliminate contaminated samples effectively and to retain as many normal samples as possible. In contrast, LogPurge achieves more precise elimination of abnormal samples with minimal loss of normal ones, benefiting from its enhanced semantic understanding and optimized purification strategy. By integrating clustering with domain rules to strengthen the LLM's discriminative capability, LogPurge enables accurate removal of highly contaminated samples and refined purification in low-contamination regions, addressing the limitations of baseline methods and compensating for the lack of semantic understanding in traditional approaches.

\subsection{RQ2:Effectiveness of Data Purification on Lightweight Anomaly Detection}\label{RQ2:Effectiveness of Data Purification on Lightweight Anomaly Detection}
\begin{table*}[t]
\centering
\footnotesize
\caption{Anomaly detection performance of all baselines. Only baselines with contaminated training are considered for metric bolding.}
\resizebox{\textwidth}{!}{
\begin{tabular}{c lcccccccccccc}
\toprule
\multirow[c]{2.5}{*}{\textbf{Baseline Type}} & 
\textbf{Dataset} &
\multicolumn{3}{c}{\textbf{BGL}} &
\multicolumn{3}{c}{\textbf{Liberty}} &
\multicolumn{3}{c}{\textbf{Industry}} \\
\cmidrule(lr){3-5} \cmidrule(lr){6-8} \cmidrule(lr){9-11}
&
\textbf{Metric} & P (\%) & R (\%) & F-1 (\%)  &
P (\%) & R (\%) & F-1 (\%) &
P (\%) & R (\%) & F-1 (\%) \\
\midrule
\multirow{1}{*}{\textbf{Oracle}} & 
\textbf{Oracle} &
92.47 & 95.07 & 93.75 & 
99.33 & 97.75 & 98.53 & 
98.86 & 99.69 & 99.27 & \\
\midrule
\multirow{2}{*}{\shortstack{\textbf{Machine Learning}\\ \textbf{Methods}}} & 
PCA~\cite{xu2009largescale} &
28.30 & 99.91 & 44.11 & 
57.14 & 21.07 & 30.79 & 
21.36 & 16.42 & 18.57 & \\
&
IsolationForest\cite{liu2008isolation} &
\textbf{94.88} & 32.13 & 48.00 & 
67.86 & 4.11 & 7.75 & 
25.64 & 1.52 & 2.87 & \\
\midrule
\multirow{3}{*}{\shortstack{\textbf{Lightweight Anomaly}\\ \textbf{Detection Methods}}} & 
DeepLog\cite{du2017deeplog} &
44.65 & 58.69 & 50.71 & 
28.62 & 22.52 & 25.21 & 
5.19 & 1.71 & 2.57 & \\
&
PLELog\cite{yang2021plelog} &
33.28 & 35.68 & 34.44 & 
32.52 & 27.33 & 29.70 & 
6.92 & 3.11 & 4.29 & \\
&
LogBERT\cite{guo2021logbert} &
88.43 & 49.32 & 63.32 & 
51.17 & 16.37 & 24.80 & 
\textbf{67.90} & 2.87 & 5.50 & \\
\midrule
\multirow{3}{*}{\shortstack{\textbf{Sample Selection}\\ \textbf{Methods}}} & 
FINE\cite{kim2021fine} &
47.90 & 98.73 & 64.51 & 
73.52 & 34.28 & 46.76 & 
10.03 & 5.65 & 7.23 & \\
&
PLUTO\cite{ma2024pluto} &
47.99 & 92.72 & 63.25 & 
72.57 & 37.30 & 49.27 & 
15.03 & 5.84 & 8.41 & \\
&
LogCleaner\cite{zhang2024reducing} &
38.32 & 69.93 & 49.51 & 
69.46 & 38.80 & 49.79 & 
27.05 & 33.97 & 28.44 & \\
\midrule
\multirow{3}{*}{\shortstack{\textbf{LLM-based Anomaly}\\ \textbf{Detection Methods}}} & 
GPT-4o\cite{achiam2023gpt} & 
37.44 & 81.52 & 51.31 & 
46.80 & 87.55 & 61.00 & 
32.46 & 97.15 & 48.66 & \\
&
LogPrompt\cite{liu2024logprompt} &
19.96 & 40.32 & 26.70 &
45.02 & 72.63 & 55.59 & 
49.96 & 96.89 & 65.92 & \\
&
LogRules\cite{huang2025logrules} &
33.17 & \textbf{99.93} & 49.81 &
- & - & - & 
- & - & - & \\

\midrule
\multirow{1}{*}{\textbf{Proposed}} &
\textbf{LogPurge} &
84.12 & 73.76 & \textbf{78.60} & 
\textbf{86.42} & \textbf{95.45} & \textbf{90.71} & 
50.13 & \textbf{99.65} & \textbf{71.02} & \\
\bottomrule
\end{tabular}
}
 \label{exp_anomaly}
\end{table*}
Table ~\ref{exp_anomaly} presents the performance of logarithmic anomaly detection as a downstream task in the BGL, Liberty and Industry datasets. We assume the existence of an oracle, trained on the LogBERT algorithm, that can perfectly distinguish between normal and abnormal log sequences. As expected, the oracle achieves the optimal results, with F1 scores above 98\% across all datasets, indicating that the anomaly detection algorithm performs remarkably well when there is no noise contamination.

For models trained on contaminated data, overall performance drops significantly. In the BGL dataset, traditional methods such as PCA and IsolationForest show extremely unbalanced precision-recall trade-offs (for example, PCA achieves P = 28. 30\% vs R = 99. 91\%), reflecting their poor adaptability to noisy logs. Even deep learning-based baselines such as DeepLog, PLELog, and LogBERT show severe degradation, with F1-scores ranging only from 49\% to 64\%. This indicates that when the training data contains heavy contamination, these models tend to overfit abnormal patterns or misclassify normal sequences as anomalies.

Similar patterns appear in the Liberty dataset, where all baselines exhibit low F1-scores (< 50\%). On the Industry dataset, which contains more realistic and heterogeneous log distributions, performance further deteriorates-most methods yield F1-scores below 10\%, suggesting extreme sensitivity to abnormal and domain-shifted training samples.

In contrast, LogPurge demonstrates consistently superior results, achieving F1-scores of 78.60\%, 90.71\%, and 71.02\% on BGL, Liberty, and Industry, respectively. Compared with the second-best performers (LogBERT on BGL, LogCleaner on Liberty, and LogCleaner on Industry), LogPurge achieves 24.1\%, 82.19\%, and 149.72\% relative improvements. These consistent gains validate the effectiveness of LogPurge's sample purification strategy, which balances the trade-off between pollution rate reduction and sample retention, leading to more robust model generalization in noisy environments.

Overall, these results confirm that (1) performance degradation under polluted training is substantial across all baselines, (2) excessive sample filtering can negatively impact model robustness, and (3) our proposed LogPurge framework achieves the best balance between data cleanliness and representativeness, significantly enhancing downstream anomaly detection performance in real-world settings.

\subsection{RQ3:Cost-Effectiveness of LLM-based Purification for Anomaly Detection}\label{RQ3:Cost-Effectiveness of LLM-based Purification for Anomaly Detection}
While LogPurge demonstrates that leveraging LLMs for data quality improvement can substantially enhance lightweight anomaly detection models, it also raises a natural question: \textit{Why not directly employ LLMs for log anomaly detection?}

To explore this question, we investigated several representative LLM-based log anomaly detection methods and compared their performance against directly using LLMs for detection. LogPrompt defines a general chain-of-thought (CoT) reasoning process for anomaly detection but lacks detailed analysis tailored to specific log datasets. LogRules, in contrast, incorporates template-level domain rules for specific datasets, yet these rules exhibit poor generalization-when logs deviate from predefined templates, detection accuracy degrades sharply.

We conducted comparative experiments from both effectiveness and efficiency perspectives, as summarized in Table~\ref{exp_anomaly}. Since LogRules provides rule sets only for the BGL dataset, its results are reported exclusively on BGL. Our findings reveal that directly applying GPT-4o achieves 48-61\% detection accuracy across three datasets but suffers from a notably high false-positive rate, indicating that LLMs without domain guidance fail to fully capture log semantics. LogPrompt, though more general, shows inconsistent performance-effective on the Industry dataset but with degraded results on two public benchmarks-suggesting that a single generic reasoning framework cannot adapt to diverse log structures. LogRules achieves the highest recall but at the cost of lower precision, reflecting its tendency to over-identify anomalies once rule bias is introduced.

As reported by Ma~\emph{et al.} ~\cite{ma2025practitioners}, over half of professionals engaged in log anomaly detection expect anomalies to be identified within 5 seconds per sequence(s/seq). However, even the fastest GPT-based approach reported in ~\cite{cui2025logeval} does not satisfy this requirement when detecting log anomaly sequences. According to our experiment, the inference times for the LLM-based methods are 11.4215 s/seq for GPT-4o, 11.9654 s/seq for LogPrompt, and 12.1342 s/seq for LogRules. In contrast, LogPurge, by purifying log data and enabling lightweight deep models for detection, achieves an inference time of only 0.0096 s/seq. This means LogPurge achieves up to \textbf{1194×} faster inference than the LLM-based methods. This difference demonstrates that purification effectively bridges the gap between LLM reasoning capability and real-time industrial demands.

\subsection{RQ4:Effectiveness of LogPurge Components}\label{RQ4:Effectiveness of LogPurge Components}
\subsubsection{Ablation Study}
\begin{figure}[htbp]
    \centering
    \includegraphics[width=\linewidth]{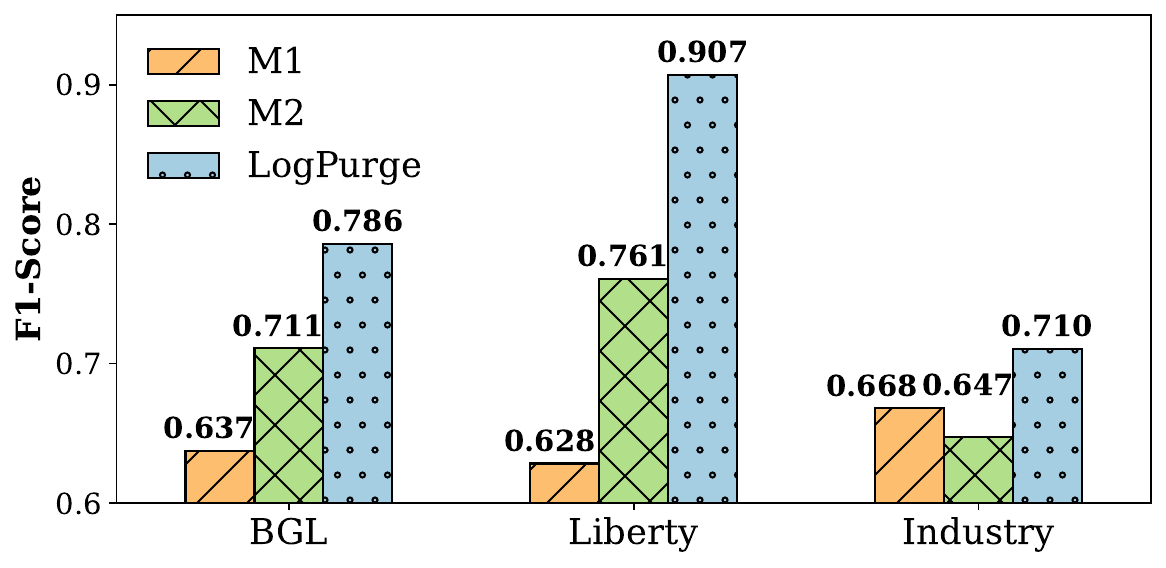}
    \caption{Ablation study of \textsc{LogPurge}}
    \label{fig:ablation study1}
\end{figure}

To evaluate the contributions of the main components in \textsc{LogPurge}, we create the following variants, reporting their results in Figure~\ref{fig:ablation study1}:
\begin{itemize}
    \item Regional Subdivision Purification Ablation(M1): We keep rule-enhanced purification as a high-contamination filter method, after that we directly use the subset as our final train subset to train a lightweight log anomaly detection method, skipping the process of subdivision and purification of low-contamination regions. 
    \item Iterative Refinement Ablation(M2): We remove the Iterative Refinement component to show how effectiveness impacts the LLM performance in log purification task. We use the original GPT-4o as LLM evaluator to classify high- or low- contamination areas without any rules.
\end{itemize}

Figure~\ref{fig:ablation study1} highlights the importance of the two core components in LogPurge, regional subdivision purification, and iterative refinement, in the log purification task. The complete LogPurge framework achieves the best performance across all datasets, with F1-scores of 0.786 on BGL, 0.907 on Liberty, and 0.710 on Industry. The improvement on the Liberty dataset is particularly remarkable, where LogPurge outperforms M1 (0.628) and M2 (0.761) by 44.43\% and 19.18\%, respectively. These results indicate that regional subdivision and further purification of low-contamination areas effectively enhance data quality, while iterative refinement improves the stability and precision of LLM-based contamination evaluation. The integration of both modules enables LogPurge to achieve superior generalization and robustness across diverse scenarios.

\subsubsection{Region Partitioning Effectiveness Study}
In the Regional Subdivision Purification stage, we convert the task of removing contamination from low-contamination regions into multiple high-contamination removal problems through effective region subdivision. Proper clustering ensures anomalous data are grouped into distinct clusters, improving subset optimization efficiency. Conversely, poor clustering may hinder reliable downstream results.
We evaluate the clustering quality using the \textit{homogeneity score}:
\[
H = \frac{1}{N}\sum_{i=1}^{N}\mathbb{I}(\text{cluster}(i)=\text{label}(i))
\]
where \(N\) is the number of log sequences, \(\text{cluster}(i)\) denotes the cluster assignment of the \(i\)-th sequence, \(\text{label}(i)\) its true class (normal/anomalous), and \(\mathbb{I}\) is the indicator function.

\begin{figure}[htbp]
    \centering
    \includegraphics[width=\linewidth]{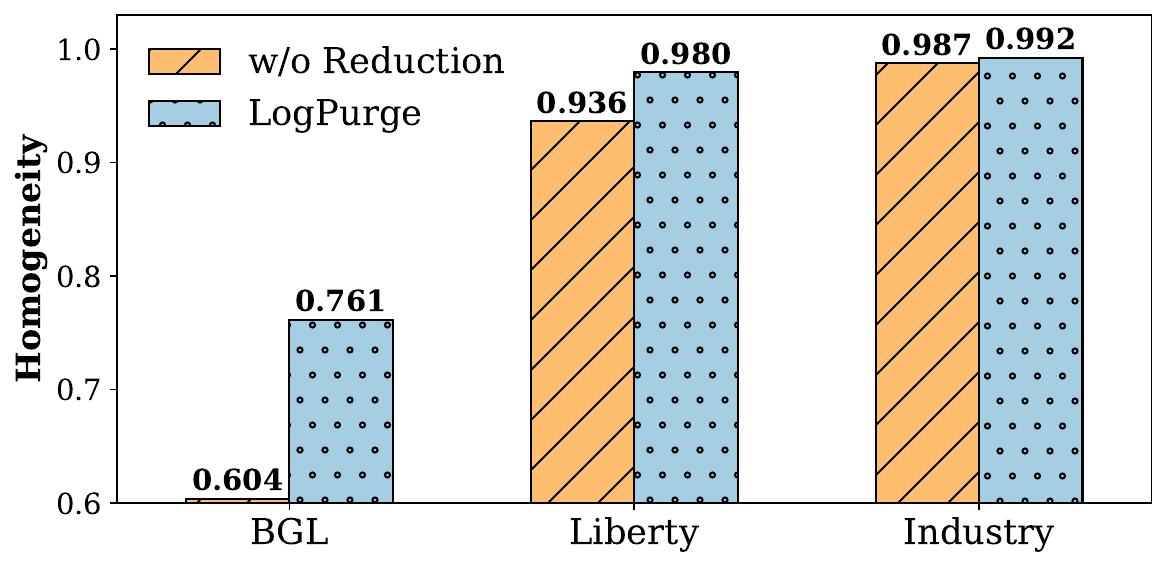}
    \caption{Region Partitioning Effectiveness study of \textsc{LogPurge}}
    \label{fig:ablation study2}
\end{figure}

LogPurge employs a dimensionality reduction–based clustering strategy to improve the homogeneity of anomaly clusters. We compare our approach with conventional clustering methods without dimensionality reduction. As shown in Figure~\ref{fig:ablation study2}, our method produces more homogeneous anomaly clusters across three benchmark datasets, thereby providing a stronger foundation for downstream anomaly detection. Moreover, the advantage becomes particularly evident in challenging scenarios (e.g., a 25.99\% improvement on BGL), where conventional methods struggle to separate noisy or overlapping regions. This improvement stems from the dimensionality reduction step, which filters out noise and redundant information, allowing clusters to form in a cleaner latent space and enhancing the consistency within each cluster.
\subsection{Additional Study}
\subsubsection{Foundation Model Study}
\label{sec:Foundation Model Study}
We further conduct a study on different base LLMs to evaluate the robustness and practicality of LogPurge. Specifically, we investigate whether the framework remains effective when replacing its underlying LLM backbone and explore the transition from closed-source to open-source models to enhance deployability and data privacy. We replace the default GPT-4o with the powerful open-source model Qwen3-235B-A22B~\cite{yang2025qwen3}(denoted as LogPurge-Qwen), while keeping all other settings unchanged. The performance is assessed using SP, CRR, and downstream F1 metrics to reveal how variations in base models influence purification quality and anomaly detection accuracy.
\begin{figure}[htbp]
    \centering
\includegraphics[width=\linewidth]{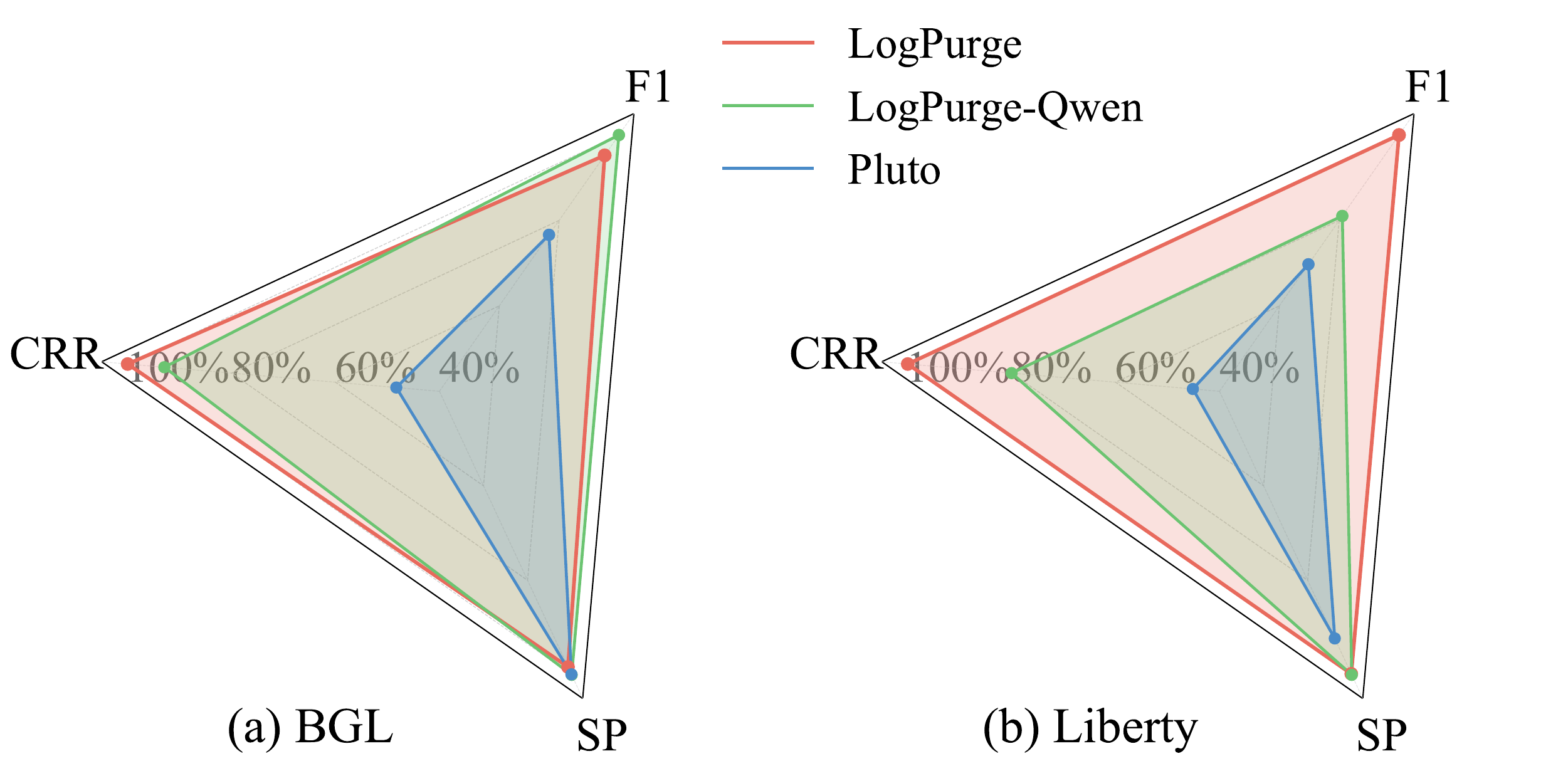}
    \caption{Foundation Model Study of \textsc{LogPurge}. 
    Each metric is normalized such that the best-performing value among all methods is scaled to 100\%.}
    \label{fig:foundation model study}
\end{figure}

Figure~\ref{fig:foundation model study}(a) and Figure~\ref{fig:foundation model study}(b) present the experimental results. After replacing the backbone model, LogPurge-Qwen still demonstrates strong performance, achieving significantly higher scores than Pluto across all three metrics, which highlights its solid tolerance to base-model variations.
Interestingly, we observe that on the BGL dataset, LogPurge-Qwen even surpasses the original LogPurge in both F1 and SP, further confirming the robustness and generalization capability of our framework.

\subsubsection{Hyperparameter Sensitivity Analysis}
To further assess the stability and sensitivity of our framework, we conduct a series of hyperparameter experiments. Specifically, we examine four key parameters: $n$, $r_{\min}$, $M$, and \textit{percentile}. The parameter $n$ controls the number of iterations in the rule enhancement process, thereby determining the degree of refinement in the purification rules. The parameters $r_{\min}$ and $M$ jointly influence the selection of representative sequences within each cluster, balancing the trade-off between diversity and representativeness of training samples. Finally, the \textit{percentile} parameter in clustering distinguishes low-contamination regions from highly contaminated ones, guiding the adaptive separation of anomalous clusters. By systematically varying these parameters, we analyze their effects on the purification quality.

\begin{figure}[htbp]
    \centering
    \includegraphics[width=\linewidth]{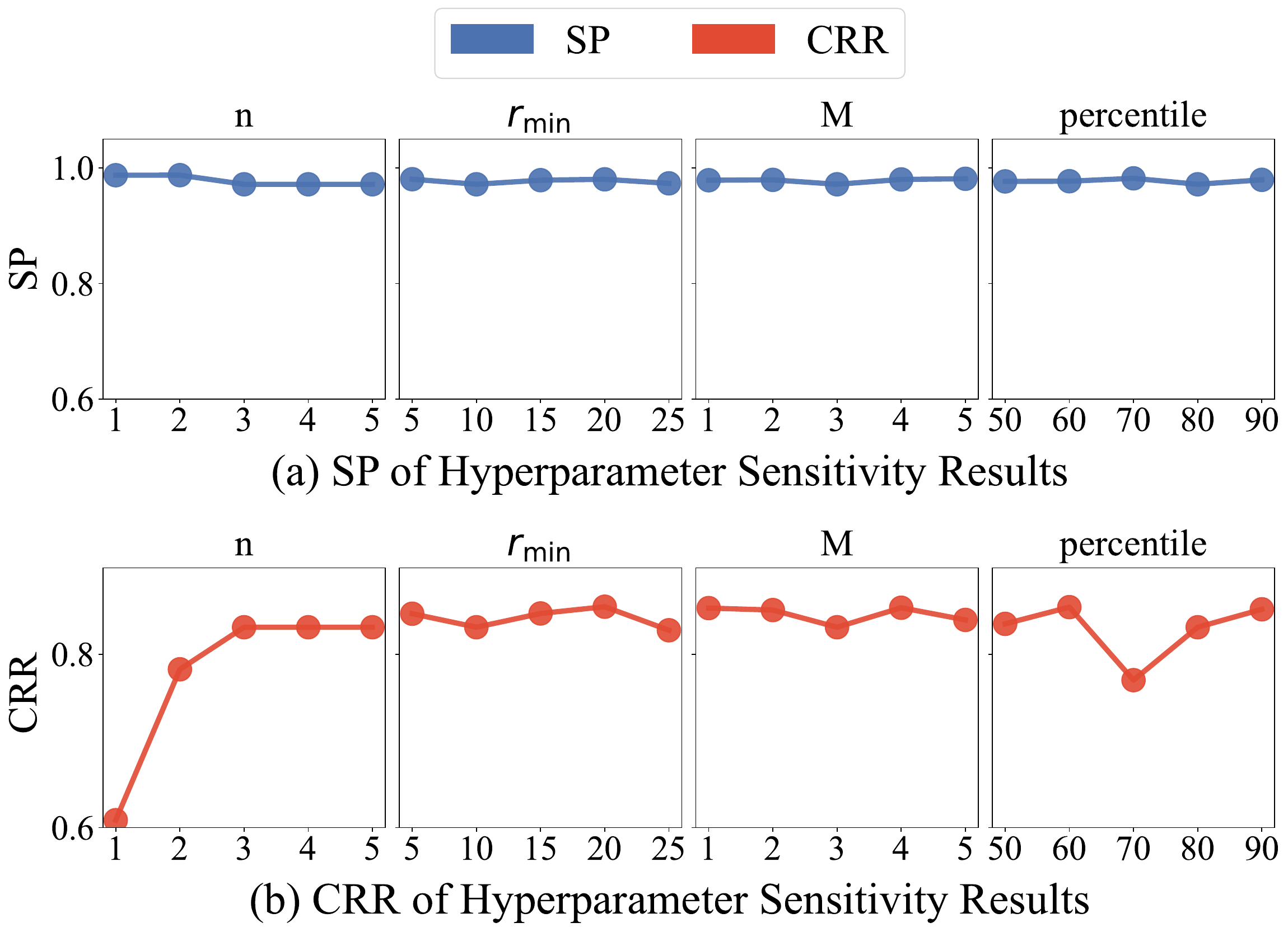}
    \caption{Hyperparameter Sensitivity Analysis of \textsc{LogPurge} on the BGL Dataset.}
    \label{fig:hyperparameter_exp}
\end{figure}

As shown in Figure~\ref{fig:hyperparameter_exp}, the performance with respect to the iteration number $n$ first improves slightly and then stabilizes, indicating that as the number of iterations increases, the induced rules gradually converge and the purification process reaches a steady state, where further iterations have negligible impact on the results. For the other three hyperparameters, including $r_{\min}$, $M$, and \textit{percentile}, both SP and CRR exhibit consistently stable performance across a wide range of values, with all results remaining above 0.8 (except for the CRR at the 70th percentile) and showing only minor fluctuations. These findings demonstrate that our framework is robust to hyperparameter variations, and its rule-enhanced purification mechanism generalizes well without the need for fine-grained tuning.

\section{Related Work}
\textbf{Log Anomaly Detection Strategies:}
In log anomaly detection, researchers have developed various strategies to identify abnormal system behaviors. Traditional machine learning approaches mainly rely on feature engineering and statistical models, such as  PCA~\cite{xu2009largescale}, IsolationForest~\cite{liu2008isolation}, and LOF~\cite{breunig2000lof}. These methods perform well on structured or low-dimensional features but struggle to capture complex temporal and semantic dependencies in log data.

Recently, deep learning has become the dominant paradigm for log anomaly detection. These approaches~\cite{du2017deeplog, yang2021plelog, guo2021logbert, meng2019loganomaly, zhang2022cat, li2022unsupervised, wang2021multi} typically treat log event sequences as time series or natural language sequences and employ deep neural networks to model contextual relationships between events, enabling fine-grained detection of anomalies.  For instance, LogAnomaly~\cite{meng2019loganomaly} applies LSTM with semantic embeddings to detect sequential and quantitative anomalies, while CAT~\cite{zhang2022cat} employs a content-aware Transformer to encode event sequences and detects anomalies via a one-class latent representation and sequence reconstruction loss.

Although deep models capture semantics and behavioral patterns more effectively, they typically require large, clean datasets. In practice, noisy logs often lead to overfitting and degraded performance, posing a key challenge for real-world applications.

\textbf{Sample Selection in Learning with Noisy Data:}
In learning tasks with noisy labels, sample selection aims to identify reliable clean samples to enhance model robustness and generalization. Although Pluto~\cite{ma2024pluto} is currently the only algorithm specifically designed for log data purification, general sample selection strategies are widely used in noisy-label learning~\cite{song2022learning}.
Multi-Network Learning ~\cite{jiang2018mentornet, han2018co, wei2020combating} selects clean samples via collaboration among multiple networks using the small-loss criterion. Multi-Round Learning~\cite{shen2019learning, huang2019o2u, song2021robust} iteratively refines the training set by selecting reliable samples in each round. Hybrid approaches~\cite{song2019selfie, zhou2020robust} combine sample selection with semi-supervised learning or loss correction techniques to exploit the potential value of unselected samples.
However, these methods largely overlook the unique structure of noisy log data, where anomalies are often concentrated and exhibit strong temporal and semantic dependencies, making conventional selection algorithms ineffective at separating contaminated anomalies from normal samples.

Even log-specific event-based selection methods~\cite{zhang2024reducing} face limitations in contaminated log scenarios. Although they remove duplicative- and anti-events to produce a compact training set, their event-level supervision overlooks sequence-level dependencies essential for log semantics, making accurate diagnosis under noisy conditions challenging.
\section{Limitation and Future Exploration}
Since \textsc{LogPurge} is an LLM-driven approach, its effectiveness is highly dependent on the underlying LLM's reasoning and comprehension capabilities. Although our framework has demonstrated robust performance across multiple LLMs in the previous experiments(Section ~\ref{sec:Foundation Model Study}), its accuracy and generalization still rely on the inherent quality of the model’s semantic understanding. In future work, we plan to collect high-quality, domain-specific log data and leverage the fault labels and explanations generated by GPT-4o to fine-tune small-scale LLMs. This adaptation aims to enhance the model’s ability to capture fine-grained log semantics, expand the applicability of \textsc{LogPurge}, and further reduce the computational and inference costs associated with large models.
\section{Conclusion}
In this paper, we presented \textsc{LogPurge}, a rule-enhanced and cost-aware framework for log data purification. Unlike prior statistical or dominance-based approaches, \textsc{LogPurge} leverages LLMs augmented with system-level rules to identify and remove contaminated log clusters while preserving representative normal patterns. Through an inspire observation, \textsc{LogPurge} processes a hierarchical process by Divide-and-Conquer strategy, it progressively eliminates anomalies and produces high-quality training data for unsupervised anomaly detection. Experiments on two public and one industrial dataset demonstrate that \textsc{LogPurge} removes up to 98.74\% of anomalies while retaining 82.39\% of normal samples, yielding substantial downstream improvements in detection performance and efficiency. By transforming log purification from a purely statistical operation into a semantic reasoning process, \textsc{LogPurge} bridges the gap between LLM understanding and lightweight model training.

\bibliographystyle{ACM-Reference-Format}
\bibliography{fse-template/references}
\end{document}